\documentclass{revtex4}
\begin{document}
\title {Time and Cosmic Clock}
\author{S. C. Tiwari}
\affiliation{Institute of Natural Philosophy\\
C/o 1 Kusum Kutir, Mahamanapuri\\
Varanasi 221005, India\\}
\email{vns_sctiwari@yahoo.com}
\maketitle

Reappraisal of the standard models (SM) of fundamental forces 
(renormalizable gauge theory) and the large scale structure of the universe 
(i.e. hot big-bang scenario) has been motivated by observations like high 
precision electro-weak measurements, quark-gluon plasma, neutrino oscillations,
 supernova Hubble diagrams and fluctuations in the cosmic microwave 
background, as well as hitherto unresolved conceptual problems. A strong 
belief has emerged that the classical notion of space-time (a la general 
relativity or GR) at Planck epoch in the early universe breaks down, and 
that quantum mechanics (QM) may not be valid near black holes. Many 
speculations have been inspired with the aim of developing a theory of 
everything. Though reluctantly, anthropic principle has also been invoked. 
A nice discussion by Dine with emphasis on effective field theories \cite{1}, 
Hogan's candid review in the context of the anthropic principle \cite{2}, and 
first two chapters of \cite{3} on the foundational issues in the unified theories 
are referred for details. In this essay, I propose to present a radical yet 
simple approach founded on the space-time interaction hypothesis (STIH) \cite{4}. 

There are numerous extrapolations beyond the SM, and it is not clear at 
present which one is preferred by nature. I have argued that in the progress 
of science at some stage a simple and pure speculative idea (fundamental 
postulate or FP) resolves outstanding problems and gives a new direction to 
the scientific thought \cite{5}. Quantized space-time is one such idea holding 
hope for 'a new revolution' \cite{6}, but being based on quantizing the gravity I 
do not share optimism for its present form. Treating QM and relativity as FPs, 
ia fresh look on their foundations is suggested. The well known problems 
extensively discussed in the literature are: physical reality of the 
wave-function and incompleteness arguments of EPR in QM, the provisional 
role of the energy-momentum tensor term in the Einstein field equation 
recognised by Einstein himself \cite{7}, and the role of physical experiences 
in GR \cite{8}. Instead of them, one single concept that could be the source of 
many problems is that of time. However, I will not discuss the questions 
like unambiguous definition of time operator and tunneling time in QM, and 
the difficulties arising from the difference between the time used in 
QM and GR; the attention will be focussed on the new concept of time \cite{9}.

Special relativity (SR) is claimed to have discarded Newton's absolute time. 
Is it true? We follow Einstein's own presentations to examine this claim. 
Motion is a pre-supposed concept, and assuming uniform motion intertial 
frame (K) is postulated. According to him, if the time of an event is $K^1$ 
is the same as the time in $K$ then it is absolute. Instantaneous  signals 
and the clocks unaffected by motion are needed to make this absolute time 
physical. Since no such signals/clocks exist time is not absolute. 
Postulating the constancy of the velocity of light in vacuum, a scheme to 
measure time is proposed. In a frame $K$, a family of clocks $\{U_n\}$ relatively 
at rest are placed at all points of space. The light signal from one of the 
clocks $U_k$ sent at the instant $t_k$ travels in vacuum to $U_l$
 placed at a distance 
$d_{lk}$ which is now set to indicate the time $t_l = t_k + dl_k/c$. 
The set of inertial 
frames $\{K_n\}$ is regulated in this way. The simultaneity of events is related by 
clock synchronisations, and the time specification depends on the space of 
reference. Reading Newton \cite{10} it becomes clear that he is precise in making 
the difference between absolute and relative for space, time and motion. 
Einstein's concern is with the measurement of relative time. What is the meaning 
of the instant tl recorded by the clock $U_l$? Einstein is silent on this, and 
later expositions/criticisms have not thrown any light on this question \cite{5,9}. 
Reichenbach tries to define it using a signal which can be marked distinctly; 
ascertaining the marks a temporal sequence is established. He is unable to propose 
any physical signal with this property, and memory, logical decision-making and 
instantaneous response are the attributes assumed for the hypothetical detection 
device. Bunge identifies 'protophysics' in SR, and terms Newtonian absolute time 
as metaphorical but fails to free SR from what he terms misinterpretations. 
Recalling Cajori's remark that absolute time is a metaphysical concept for Newton, 
we conclude that SR does not address the problem of the nature of time. 

The problems gets worse in GR; shown in a dramatic way by Godel's rotating universe. 
Time-like and null vectors are used to get time directionality, but it is found that 
no time coordinate exists at each space-time point that increases as one moves in a 
positive time-like direction. Einstein dismisses Godel universe as unphysical but 
accepts the difficulty of time ordering in relativity \cite{10}. In cosmology, it is 
possible to make mathematically precise definition of homogeneity and isotropy 
using isometry of the metric and congruence of time-like curves \cite{12}, but 
physically it corresponds to usual world-view of $3+1$ dimensions. Assumption of 
local comoving frames and ideal clocks not only fails at the big-bang singularity 
and near massive bodies, it is not justified for a theory purporting to be a 
fundamental theory \cite{9}. Finally I refer to a new paradox pointed out in 1988 \cite{9}: 
The most significant direct evidence for SR comes from the observed enhanced 
lifetime of a fast moving unstable particle (e.g. muon). Decay is an irreversible 
process, and relativistic time dilation arises in a time-reversal symmetric 
kinematics. Equality of unidirectional lapse of time (lifetime) with time 
dilation leads to a paradoxial situation. 

Departing from the contemporary trend accepting relativistic world-view, 
absolute nature of time is proposed based on STIH stated as: Time is the 
primeval cause transforming the source (or primordial) space into the 
physical space $(S \otimes P)$.

Source space is inert, time ceaselessly transforms this into active or manifest 
universe i.e. the pre-universe state or beyond it is that called the source space.
This transformation represents the flow of time, and all changes in the universe 
take place with reference to this time, hence it is absolute. Time is 
unidirectional due to irreversible process $S \otimes P$. In any physical measurement, 
the direct observations correspond to the counting of spatial lengths and spatial 
relations which are related to the physical quantities based on a calibration 
procedure. Thus irreducible part of any measurement is counting of a unit of 
length, and for observed three dimentional space a reasonable measure is its 
volume. A measurable duration of elementary time interval for successive action 
transforming $S \otimes P$ can be equivalently used. Let the volume $V_e$ is transformed 
in a duration of time interval $t$. Simplest choice is to assume creation of equal 
volume $V_e$ in each time step $t$. Noting the importance of the concept of energy, 
we may endow the source space with fundamental energy, and express $V_e$ and $t$ in 
the units of energy. Assuming Planck constant as a unit conversion factor, 
the energy in volume $V_e$ is $h/t$, and energy density $\rho$

\begin{equation}
\rho V_e  =  h/\tau
\end{equation}

The volume of the universe at time $T = N\tau$, (N is an integer) is
\begin{equation}
V(T)  =  V_eT/\tau	
\end{equation}

For sufficiently large $N$, a continuous variable $t$ may be used instead of $T$.

To build a simple model, we assume spherical (spatial) universe. Incremental 
radius $\Delta R$ from $(N-1)\tau$ to $N\tau$ is given by

\begin{equation}
\Delta R  =  R(N\tau) - R[(N-1)\tau]
\end{equation}

For large N, following expressions are easily obtained

\begin{equation}
\frac{dR}{dt}  =  \frac{1}{3} \lambda t^{-2/3}
\end{equation}
\begin{equation}
\frac{d^2R}{dt^2}  =  - \frac{2}{9} \lambda t^{-5/3}
\end{equation}
where the constant
\begin{equation}
\lambda  = \left(\frac{3V_e}{4\pi\tau}\right)^{1/3}
\end{equation}

Interesting physical consequences follow from this model \cite{3} assuming $t$ equal to 
Planck time ($\approx 5.5\times 10-44 sec$), and imposing the condition that 
at any time the 
maximum velocity is limited by the cosmic expansion rate Eq.(4). Taking input data, 
age of the universe \cite{13} 12.5 billion yrs and velocity of light equal to $dR/dt$ 
at present, the size of the universe in the first instant is calculated to 
be $\approx 108$ cm, in contrast to Planck length in big-bang model. Equation (1) 
gives the value of the energy density, and again treating $c$ as a unit conversion 
factor, we can express it in equivalent mass density ($\approx 5\times 10-30 gm/cm3$). 
Equating the acceleration due to gravity of the total energy of the universe on itself of 
Newtonian gravition by Eq.(5), the value of gravitational constant $G$ can be determined. 
In this case, both age of the universe and r can be derived assuming the known 
values of $G$ and $c$. Evidently, in this model

\begin{equation}
G(t) \propto t^2
\end{equation}
\begin{equation}
c(t)  =  \frac{dR}{dt} \propto t^{-2/3}
\end{equation}

Time variation of physical constants has been discussed e.g. by Dirac and now 
in the context of compactification of the extra dimensions in Kaluza-Klein/superstrings. 
As an alternative to inflationary model \cite{14}, time varying light velocity cosmologies 
have recently been proposed \cite{15}. Unlike them, our approach does not give a 
fundamental 
role to the velocity of light. In fact, a cosmic clock gives an absolute measure of 
time, and cosmic expansion rate regulates the physical processes. Using empirical 
observations on time variations of $c$ and $G$, stringent limits on the unknown parameters 
in our model like $r$ and age of the universe can be obtained.

Radial expansion and internal oscillations/rotations constitute basic dynamical 
processes. Elementary entities are proposed to be the spatio-temporal bounded 
structures with characteristic sizes $DR$ at different epoch create many such 
species: at 1 sec $DR \approx 10-21$ cm and at 1010 yrs it is $\approx 10-33$ cm. Lumps of 
aggregates of these elementary species is matter in this scenario. 
A significant development in STIH is envisaged with the proposition that 
electronic charge is a manifestation of fractional spin $e2/c$ of a 2+1 dimensional 
model of electron \cite{3}, see also discussion in \cite{16}. In contrast to the question 
asking how to get 4D from higher dimensions, we find 2D structures in the 3D space. 
The fractional spin has different values for different $DR$ created at different moments 
during the evolution of the universe, therefore the spatial location of the planet 
earth and its age determine the values of the coupling constants (like fine structure) 
and the existence of life here, however species with different coupling constants may 
have formed varied structures elsewhere (e.g. in the extra-galactic universe).

Incorporating interaction (spin origin) and using statistical mechanics to 
understand the structure of the universe are the problems deserving attention, 
though the slow progress in STIH has been mainly due to efforts of the author 
in isolation. A measure manifold for the cellular structure of space-time in 
unimodular gravity \cite{17} appears a promising candidate for enunciating an 
extremum principle.

This work is not funded by any agency. Library facility at Banaras Hindu University 
is acknowledged.

\end{document}